\documentclass[journal,10pt,letterpaper,final,twocolumn]{IEEEtran}%
\usepackage{amsmath}
\usepackage{amsfonts}
\usepackage{cite}
\usepackage{amssymb}
\usepackage{mathrsfs}
\usepackage{graphicx}
\usepackage[usenames,dvipsnames]{color}
\usepackage{epstopdf}
\usepackage[all]{xy}
\usepackage[]{mcode}
\usepackage{algorithm}
\usepackage{algorithmic}
\usepackage{subcaption}
\usepackage{array}

\usepackage{optidef}

\providecommand{\U}[1]{\protect\rule{.1in}{.1in}}
\pdfoutput=1
\providecommand{\U}[1]{\protect\rule{.1in}{.1in}}

\newcommand{\qed}{\nobreak \ifvmode \relax \else
      \ifdim\lastskip<1.5em \hskip-\lastskip
      \hskip1.5em plus0em minus0.5em \fi \nobreak
      \vrule height0.75em width0.5em depth0.25em\fi}

\begin{document}

\title{\LARGE {Power Allocation for Relayed OFDM with Index Modulation Assisted by Artificial Neural Network}}
\author{Jiusi Zhou, \textit{Student Member, IEEE}, Shuping Dang, \textit{Member, IEEE}, Basem Shihada, \textit{Senior Member, IEEE}, and \\ Mohamed-Slim Alouini, \textit{Fellow, IEEE} 
  \thanks{The authors are with Computer, Electrical and Mathematical Science and Engineering Division, King Abdullah University of Science and Technology (KAUST), Thuwal 23955-6900, Kingdom of Saudi Arabia (e-mail: $\{$jiusi.zhou, shuping.dang, basem.shihada, slim.alouini$\}$@kaust.edu.sa).}
}

\maketitle

\begin{abstract}
In this letter, we propose a power allocation scheme for relayed orthogonal frequency division multiplexing with index modulation (OFDM-IM) systems. The proposed power allocation scheme replies on artificial neural network (ANN) and deep learning to allocate transmit power among various subcarriers at the source and relay nodes. The objective of the power allocation scheme is to minimize the overall transmit power under a set of constraints. Without loss of generality, we assume all subcarriers at source and relay nodes are independently distributed with different statistical distribution parameters. The relay node adopts the fixed-gain amplify-and-forward (FG AF) relaying protocol. We employ the adaptive moment estimation method (Adam) to implement back-propagation learning and simulate the proposed power allocation scheme. The analytical and simulation results show that the proposed power allocation scheme is able to provide comparable performance as the optimal solution but with lower complexity.

\end{abstract}

\begin{IEEEkeywords}
Power allocation, index modulation, OFDM, amplify-and-forward relaying, artificial neural network (ANN).
\end{IEEEkeywords}

\section{Introduction}
\IEEEPARstart{T}{o} cope with rapidly increasing data demand in next-generation networks, orthogonal frequency division multiplexing with index modulation (OFDM-IM) is regarded as one of the most promising modulation candidates \cite{w1,new3,dang2020should}. Since the proposal of the canonical OFDM-IM scheme in \cite{w3}, many studies have been carried out to study the performance and optimization of OFDM-IM. In \cite{7330022}, Wen et al. rigorously proved the spectral efficiency advantage of OFDM-IM over classic OFDM from the information-theoretical perspective. Also, an enhanced OFDM-IM scheme is proposed in \cite{7929332} to provide a higher spectral efficiency and a diversity gain, which paves the way to practical implementation of OFDM-IM. Another method to raise the spectral efficiency is to introduce multiple modes, which gives the multi-mode OFDM-IM scheme \cite{7936676}. Besides spectral efficiency, improving energy efficiency and network coverage is also a direction of  communication technology development, and it is not exceptional for OFDM-IM \cite{new1}. To enhance transmit reliability and efficiency, relayed OFDM-IM was first investigated through numerical results in \cite{w2}. Currently, more studies are launched to further substantiate the superiority of relayed OFDM-IM \cite{8730295,w5,8891788,9096410,8917612}. 

Meanwhile, power allocation can be implemented in conjunction with cooperative relaying to further enhance the performance of relayed OFDM-IM. Specifically, a convex programming technique is proposed in \cite{9096410} to perform power allocation for FG AF relay assisted OFDM-IM. However, for the simplicity of simulation, most studies unify the simulation environment parameters for all subcarriers. In practice, the parameters of thermal noise are always different for different subchannels, which are related to the ambient temperature and the subcarrier bandwidth. As a result, the convex programming aided power allocation scheme proposed in \cite{9096410} might not always be applicable to the cases with different parameters for different subcarriers. 

Different from canonical convex optimization techniques, deep learning based on a well-designed artificial neural network (ANN) can be used to emulate the brain-based reasoning process and has the ability to learn from the previous samples through a back-propagation mechanism so as to improve the performance. Due to the powerful processing ability of deep learning, we design an ANN and employ deep learning to propose an efficient power allocation scheme for relayed OFDM-IM systems using subcarriers with heterogeneous statistical properties. The objective of the power allocation scheme is to minimize the overall transmit power allocated among active subcarriers at both source and relay node subject to outage and maximum transmit power constraints for relayed OFDM-IM systems. The adaptive moment estimation method (Adam) is leveraged in simulations to implement back-propagation learning.

\section{System Model}\label{sec2}
To study relayed OFDM-IM, we employ a three-node transmission scenario in this letter, in which the direct transmission link is neglected in favor of simplicity. We denote the set of $N$ subcarriers as $\mathcal{N}$, and a part of which, say $T$ subcarriers, will be activated and convey information. The transmission of these $T$ active subcarriers are sent from the source, forwarded by a fixed-gain (FG) amplify-and-forward (AF) relay, and received by the destination. In this letter, we follow the classic OFDM-IM rules stipulated in \cite{w3} for activating $T$ out of $N$ subcarriers so that a subcarrier activation pattern (SAP) is formed. The SAP can also be used to represent information, and the corresponding subset of active subcarriers is denoted as $\mathcal{T}(k)$. As a result, the length of the transmitted bit stream is $B=\lfloor \log_2 \binom{N}{T} \rfloor+T \log_2 M$, where $M$ is the order of $M$-ary phase shift keying ($M$-PSK), which is adopted as the amplitude-phase modulation scheme. For the sake of simplicity, we assume that all incoming bit streams are equiprobable. Applying $N$-point inverse fast Fourier transform (IFFT) yields the independent OFDM block written as $\mathbf{x}(k)=[x(m_1,1),x(m_2,2),\dots,x(m_N,N)]^T \in \mathcal{C}^{N\times1}$, where ${x(m_i , i)x(m_i , i)^* = 1}$ for active subcarriers, and ${x(m_i , i) = 0}$ otherwise. 

Then, with independent fading over different subcarriers, the end-to-end received signal $y(m_i,i)$ transmitted over active subcarriers becomes $y(m_i,i) = \sqrt{{P_{r,i}}{P_{t,i}}} {h_{1,i} h_{2,i} x(m_i,i)} + \sqrt{{P_{r,i}}}  {h_{2,i} w_{1,i}+w_{2,i}}$, where $P_{r,i}$ is the $i$th subcarrier transmit power of the relay node, which represents the amplification gain and is controllable at the FG AF relay node according to the statistical channel state information (CSI); $P_{t,i}$ is the $i$th subcarrier transmit power of the source; $w_{j,i}$ is the complex additive white Gaussian noise (AWGN) that is characterized by average noise power $\eta_{j,i}=\mathbb{E}\{w_{j,i}w_{j,i}^*\}$; $h_{j,i}$, $\forall{j} \in \{1,2\}$, represents the channel fading gain at the link from the source to the relay node and from the relay node to the destination respectively for subcarrier $i$. Based on the assumed fading environment, we can have the PDF and the CDF of the channel power gain $G_{j,i}=\left|h_{j,i}\right|^{2}$ to be $f_{j,i}(\xi)=\exp \left(-\xi / \mu_{j,i}\right) / \mu_{j,i}$ and $F_{j,i}(\xi)=1-\exp \left(-\xi / \mu_{j,i}\right)$, respectively, where $\mu_{j,i}$ is the average channel power gain. The independent end-to-end SNR of an arbitrary active $i$th sub-carrier is expressed as $\gamma(k,i)= \frac{P_{t,i} P_{r,i} G_{1,i} G_{2,i} }{{P_{r,i}} G_{2,i}\eta_{1,i} + \eta_{2,i}}$, $\forall~i \in \mathcal{T}(k)$.

An outage event of an OFDM transmission block happens if the end-to-end SNR of any active subcarrier is lower than a preset outage threshold $s$. Therefore, the average outage probability of the proposed relay assisted OFDM-IM with independent fading over subcarriers can be written as \cite{9096410}
\begin{equation}\label{eq44}\small
\begin{split}
P_{o}^{*}(s)&= \frac{1}{\Xi}\sum_{k=1}^{\Xi}\left\{\left[1-\prod_{i \in \mathcal{T}(k)} (1-\Phi_{i}(s))\right]\right\},
\end{split}
\end{equation}
where $\Xi=2^{\lfloor \log_2 \binom{N}{T} \rfloor}$ is the number of legitimate SAPs; $\Phi_{i}(s)$ is the outage probability for the $i$th subcarrier with a uniform outage threshold $s$. Following the derivation given in \cite{bb1}, $\Phi_{i}(s)$ is determined by $\Phi_{i}(s) =1-{2} \sqrt{\frac{s \eta_{2,i}}{\mu_{1,i} \mu_{2,i} {P_{t,i}} {P_{r,i}}}} \exp \left(-\frac{s \eta_{1,i}}{\mu_{1,i} {P_{t,i}}}\right) K_{1}\left({2}\sqrt{\frac{s \eta_{2,i}}{\mu_{1,i} \mu_{2,i} {P_{t,i}} {P_{r,i}}}}\right)$, where $K_v(\cdot)$ denotes the $v$th-order modified Bessel function of the second kind.

\section{Problem Statement}
From an energy-efficient perspective, a total transmit power minimization problem for $T$ active subcarriers is formulated as follows: 

 \begin{mini} 
 {}{\sum_{i \in \mathcal{T}(k)}\left(P_{t,i}+P_{r,i}\right)~~~~~~~~~~~~~~~~~~~~~~~~}{}{}
 \addConstraint{P_{o}^{*}(s)\leq \Psi_{{th}}}
 \addConstraint{0 \leq P_{t,i} \leq P_{t-sub}^{\max }, \forall{i} \in \mathcal{T}(k)}
 \addConstraint{0 \leq P_{r,i} \leq P_{r-sub}^{\max }, \forall{i} \in \mathcal{T}(k)}.
 \end{mini}
It is required to maintain the average outage probability below a predetermined threshold $\Psi_{{th}}$. Also, for each active subcarrier, there exist upper bounds on allocated power at the source and the relay, denoted as $P_{t-sub}^{\max}$ and $P_{r-sub}^{\max}$, due to the hardware constraints.

Observing (\ref{eq44}), we can easily notice that the average outage probability $P_{o}^{*}(s)$ is non-linear in terms of average channel power gains. This leads to the non-convexity of the formulated power allocation problem. To obtain the optimal solutions to non-convex problems, exhaustive and random searching methods could be helpful. However, they normally demand a huge amount of computational resource and time to converge to the optima, which are not suited for the real-time optimization. Recently, deep learning aided by ANN has exhibited competence to solve NP-hard optimization problems in a rapid and accurate manner \cite{8765703}. We hereby introduce deep learning and ANN to tackle the formulated power allocation problem for relayed OFDM-IM systems.

\section{Power Allocation for Relayed OFDM-IM Systems by Deep Learning and ANN}

\subsection{ANN Architecture and Data Structures}
\begin{figure}[t]
\centering
\includegraphics[width= 3.2in,angle=0]{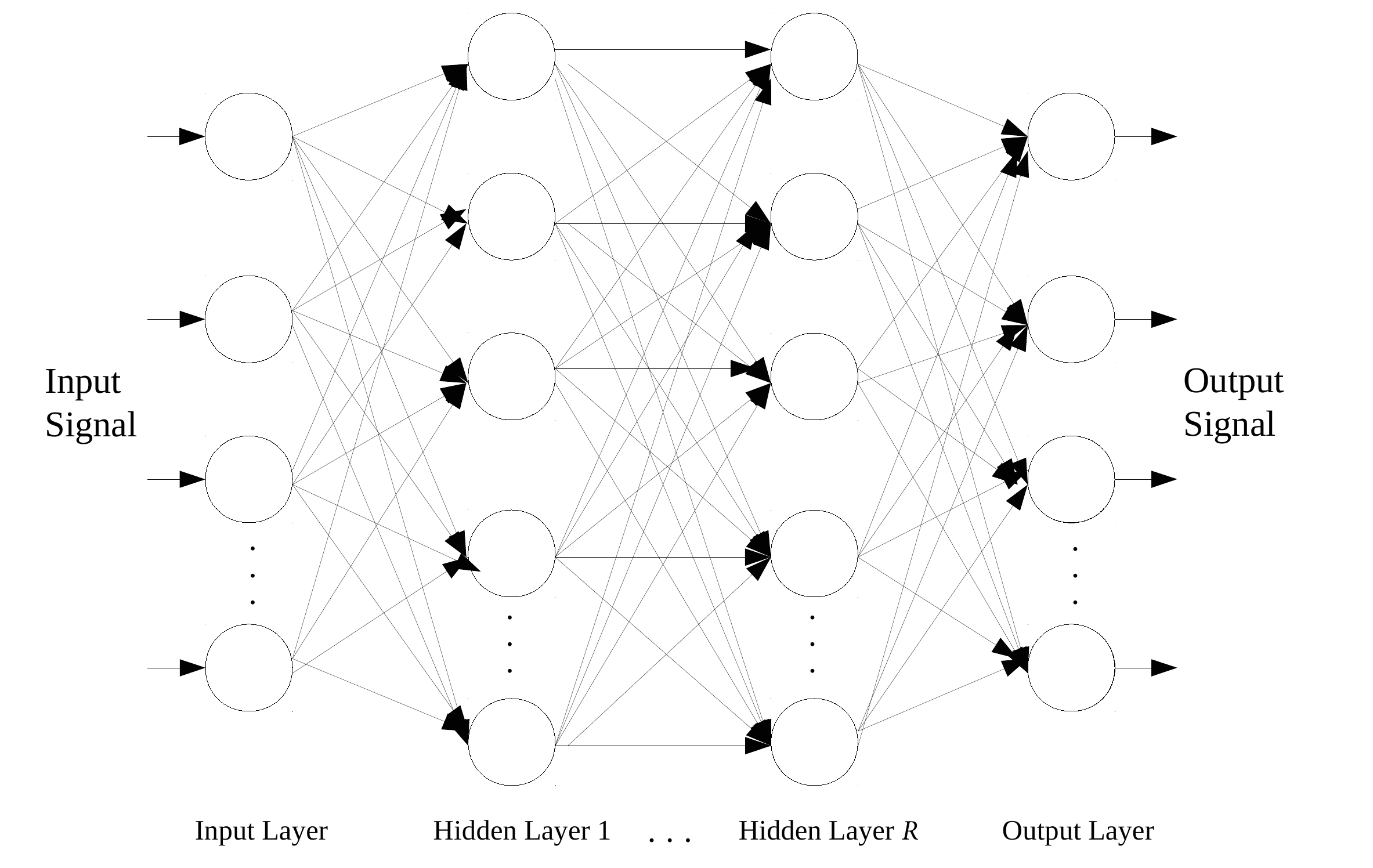}
\caption{An example of the ANN architecture.}
\label{fig:ANN}
\end{figure}

Prior to devising a specific framework of deep learning for our formulated problem, we first introduce the architecture of ANN.  As shown in Fig.~\ref{fig:ANN}, a typical ANN consists of a hierarchical structure of layers, and these layers arrange the neurons in the network. The neurons connected to the external environment form input and output layers. Adjusting the weights of links connecting neurons makes network input/output behaviors consistent with the environmental behaviors. To be specific,  the architecture design of ANN involves an input layer, $R$ hidden layers and an output layer. The amount of neuron nodes in the hidden layer is case-specific. In general, a larger number of neurons implies a higher training efficiency.

For each neuron in the hidden and output layers, it receives multiple signals from the neurons in the previous layer, calculates a new activation level, and sends it through the links connecting the neurons in the next layer. The output signals from the neurons in the output layer are organized by some post-processing techniques to yield the solution to the problem of interest. One of the key properties of an ANN is the adopted activation function, which could be the sign function, the step function, and the sigmoid function. Considering that our formulated power allocation problem is not in a binary structure, the sigmoid function $S(\cdot)$, mapping a real value to another constrained real value between 0 and 1, would suit our needs and is thus adopted to produce the ratios of allocated power for all subcarriers at the source and relay. 

Deep learning aided by a well-designed ANN can be used as a powerful tool to extract knowledge from a sufficiently large amount of empirical data. Therefore, a sub-optimal solution to the problem of interest can be produced. To utilize deep learning and ANN in an efficient manner, we first need to stipulate the data structures for the input and output layers. For $T$ active subcarriers, four categories of information, including statistical CSI and average noise power, are taken into consideration, which are $\{\mu_{1,i}\}$, $\{\mu_{2,i}\}$, $\{\eta_{1,i}\}$, and $\{\eta_{2,i}\}$. For each group of samples, we employ the exhaustive search to determine the optimal power allocation solutions, i.e., the labels corresponding to the samples. Each group of samples and the labels yielded by the exhaustive search can be written in the matrix form as

\begin{equation} \label{eq51}\small
\mathbf{V}=\left[\begin{array}{cccccc}
\mu_{1,1} & \ldots & \mu_{1,i} & \ldots & \mu_{1,T} \\
\mu_{2,1} & \ldots & \mu_{2,i} & \ldots & \mu_{2,T} \\
\eta_{1,1} & \ldots & \eta_{1,i} & \ldots & \eta_{1,T}\\
\eta_{2,1} & \ldots & \eta_{2,i} & \ldots & \eta_{2,T}
\end{array}\right] ,
\end{equation}
and
\begin{equation} \label{eq51}\small
\mathbf{U}=\left[\begin{array}{cccccc}
P_{t,1} & \ldots & P_{t,i} & \ldots & P_{t,T} \\
P_{r,1} & \ldots & P_{r,i} & \ldots & P_{r,T} 
\end{array}\right].
\end{equation}
Note that the numbers of entries of $\mathbf{V}$ and $\mathbf{U}$ are $4T$ and $2T$, respectively. That is, we need to approximate $2T$ quantities by $4T$ quantities. As a result, we construct an ANN with $4T$ and $2T$ neurons in the input and output layers to produce the sub-optimal solution based on a set of labeled training samples.

\subsection{Model Training}
With sufficient labeled training samples, the back-propagation mechanism can be employed to train the ANN model and gradually approach the optimal one with appropriate link weights. The whole training process is constituted by a number of training epochs, and we group the training samples in batches. The overall training process is carried out in a supervised manner and we begin with a loss function in this process.

For each training epoch, the optimizer needs to go through all training samples batch by batch for each time step. According to the principle of supervised machine learning, the link weights are fine-tuned based on the comparison between the output of ANN and the label. The mean squared error (MSE) between the output of ANN and the label is adopted as the loss function for the comparison. Denote $\Omega(\cdot, \{\theta_t\})$ as the ANN output matrix with the input matrix argument and the set of link weights $\{\theta_t\}$. The output of $\Omega(\cdot, \{\theta_t\})$ is a matrix having the same dimension as $\mathbf{U}$. Therefore, the loss function of the $t$th time step is explicitly given by 

\begin{equation}\small\label{lossfunctioneq}
\begin{split}
f_{t}(\mathbf{U}_{t,1},\dots,\mathbf{U}_{t,D}, \mathbf{V}_{t,1},\dots,\mathbf{V}_{t,D}, \{\theta_{t-1}\})\\
=\frac{1}{D} \sum_{d=1}^{D} \left\|\mathbf{U}_{t,d}-\Omega(\mathbf{V}_{t,d},\{\theta_{t-1}\})\right\|_F^{2},
\end{split}
\end{equation}
where $\mathbf{V}_{t,d}$ and $\mathbf{U}_{t,d}$ denote the $d$th group of samples and the corresponding label  in a single batch of the $t$th time step; $D$ is the  batch size.

Adam, proposed in \cite{aa3}, is considered as the  mainstream back-propagation method in both academia and industry and is thereby adopted in this letter. Developing from the stochastic gradient descent method, Adam computes individual learning rates for different parameters. Aiming at adapting the learning rate for each weight, the estimates of the first and second moments of the gradient are used for adaptation purposes. Similar to the stochastic gradient descent method, the adaptive algorithm of Adam is shown in Algorithm \ref{algorithm1}.

\begin{algorithm}[t!] 
 \caption{Back-propagation algorithm of Adam.}
 \begin{algorithmic}[1]
 \renewcommand{\algorithmicrequire}{\textbf{Input:}}
 \renewcommand{\algorithmicensure}{\textbf{Output:}}
 \REQUIRE $\beta_1$, $\beta_2 \in [0,1)$ (two hyper-parameters), $\epsilon$, $\Delta$ (step size)
 \ENSURE  $\theta_{t}$
  \\ \textbf{Initialization:} $m_0= 0, v_0 = 0, t = 0$;
  \WHILE {$\theta_{t}$ does not converge}
  \STATE $t \gets t+1$;
  \STATE $g_{t} \gets \nabla_{\theta_{t-1}} f_{t}$ ($\nabla_{\theta_{t-1}}$ represents the  gradient operator with respect to weights $\{\theta_{t-1}\}$);
  \STATE $m_{t} \gets \beta_{1} m_{t-1}+\left(1-\beta_{1}\right) g_{t}$, $v_{t} \gets \beta_{2} v_{t-1}+\left(1-\beta_{2}\right) g_{t}^{2}$;
  \STATE $\hat{m}_{t} \gets {m_{t}}/{(1-\beta_{1}^{t})}$, $\hat{v}_{t} \gets {v_{t}}/{(1-\beta_{2}^{t})}$;
  \STATE $\theta_{t} \gets \theta_{t-1}-\Delta {\hat{m}_{t}}/{(\sqrt{\hat{v}_{t}}+\epsilon)}$
  \ENDWHILE
  \RETURN $\theta_{t}$
 \end{algorithmic} 
 \label{algorithm1}
 \end{algorithm}


\section{ Analysis of Computational Complexity}

To clarify the motivation and reveal the technical contribution analytically, we perform the analysis of computational complexity for the power allocation scheme using ANN in this section. For an ANN with $R$ hidden layers and $\rho_r$ neurons in the $r$th layer, the data matrix in hidden layer $r$ is denoted as $\mathbf{Q}_r$; the weight matrix between $r$th layer and $(r+1)$th layer is denoted as $\mathbf{W}_r$ (Here, we refer the $0$th layer to the input layer for notational simplicity). Regarding the invoking process of the ANN model as a feedforward pass process, from layer $r$ to layer $r+1$, we can have $\mathbf{Q}_{r+1} = \mathbf{W}_{r} \mathbf{Q}_{r}$ and then apply the activation function in an entry-wise manner to have the mapping relation: $\mathbf{Q}_{r+1} \overset{S(\mathbf{Q}_{r+1})}{\longrightarrow}(0,1)^{q_{r+1}^{\mathsf{R}}\times q_{r+1}^{\mathsf{C}}}$, where $q_{r+1}^{\mathsf{R}}$ and $q_{r+1}^{\mathsf{C}}$ denote the number of rows and columns of $\mathbf{Q}_{r+1}$. In this way, we can determine the computational complexity of this operation in the $i$th layer as $\mathcal{O}(\rho_r \rho_{r+1} +\rho_{r+1} )=\mathcal{O}((\rho_r+1) \rho_{r+1} )=\mathcal{O}(\rho_r \rho_{r+1})$. Taking all $R$ hidden layers into consideration, the computational complexity of the entire feedforward propagation process is given by $\mathcal{O}(4T\rho_1+\sum_{r=1}^{R}\rho_r\rho_{r+1}+2T\rho_R)$. 

For comparison purposes, we also analyze the computational complexity of exhaustive search depending on searching accuracy $\delta$. If we simplify the function seeking the optimized solution as $\mathcal{F}(A)$ where $A$ denotes the set of object data, including $P_r$ and $P_t$. For a sub-linear convergence problem, the optimal solution $\mathcal{F}^{*}$ should satisfy $\mathcal{F}\left(A_{\kappa}\right)-\mathcal{F}^{*} \leq \frac{\varepsilon}{\sqrt{\kappa}}$, where $\kappa$ is the number of searching rounds; $\varepsilon$ is set as a constant associated with platform configurations. Let $\frac{\varepsilon}{\sqrt{\kappa}} \leq \delta$, resulting in $\mathcal{O}\left(\frac{1}{\delta^{2}}\right)$ as the computational complexity of exhaustive search. 

From the above analysis, it is obvious that the computational complexities pertaining to neural computing and exhaustive search depend on different constructions. By property adjusting the setups of an ANN, it is entirely possible that a low-complexity power allocation scheme can be provided by neural computing, which outperforms exhaustive search in terms of computational complexity.

\section{Simulation Results and Key Observations}
We investigate the effectiveness of the ANN based power allocation scheme for relayed OFDM-IM systems by comparing  with the optimal benchmark given by exhaustive search. In this study, we consider an example of ANN architecture with six hidden layers, and each layer has 128 artificial neurons. We fix the system setup and performance threshold as follow: $\Psi_{{th}} = 10^{-2}$, $s = 1$, $N=4$, and $T=2$. As a result, eight inputs are taken into consideration, including $\mu_{1,1}$, $\mu_{1,2}$, $\mu_{2,1}$, $\mu_{2,2}$, $\eta_{1,1}$, $\eta_{1,2}$, $\eta_{2,1}$, and $\eta_{2,2}$. After randomly generating 6561 groups of input training samples of these eight inputs in the range of\footnote{According to \cite{3GPP}, the channel gain coefficient is standardized to be $ \mu_{j,i} = 10^{-12.8}\lambda^{-\alpha}\tilde{\mu}$, where $\lambda$ denotes the distance between transmitter and receiver; $\alpha$ is the path loss exponent; $\tilde{\mu}$ represents a random variable abiding the Rayleigh distribution with unit mean. As for the noise power, $\eta_{j,i} = k_bT_cB$, $k_b$ denotes Boltzmann's constant $\left(1.38 \times 10^{-23}~\mathrm{J} / \mathrm{K}\right)$; $T_c$ is the thermodynamic temperature in Kelvins (290 K set in most civil application scenarios); $B$ is the bandwidth in Hz (available values for receiver bandwidth range from about 5-100 kHz). Based on these two formulas, for  general cases, $\mu_{j,i}$ and $\eta_{j,i}$ are comparable in quantity. To study the effects of different coefficients, we set the distribution ranges of both parameters as [0.5,5] without loss of generality, where the value of upper bound is 10 times as value of the lower bound.} $[0.5,5]$, we utilize exhaustive search to seek corresponding 6561 groups of labels for $P_{t,1}$, $P_{t,2}$, $P_{r,1}$, and $P_{r,2}$ and then obtain a set of labeled training samples. In addition, we set the batch size as 32 to speed up the training process. In ANN parameter setup, especially for Adam back-propagation method,  two hyper-parameters $\beta_1$ and $\beta_2$ are fixed to be 0.9 and 0.999 respectively. In addition, we let $\Delta = 10^{-4}$, $\epsilon = 10^{-8}$, set the number of training epochs to be $10^5$, and denote the set of optimized link weights as $\{\theta_*\}$.

\begin{figure}[!t]
\centering 
\includegraphics[width =2.8in]{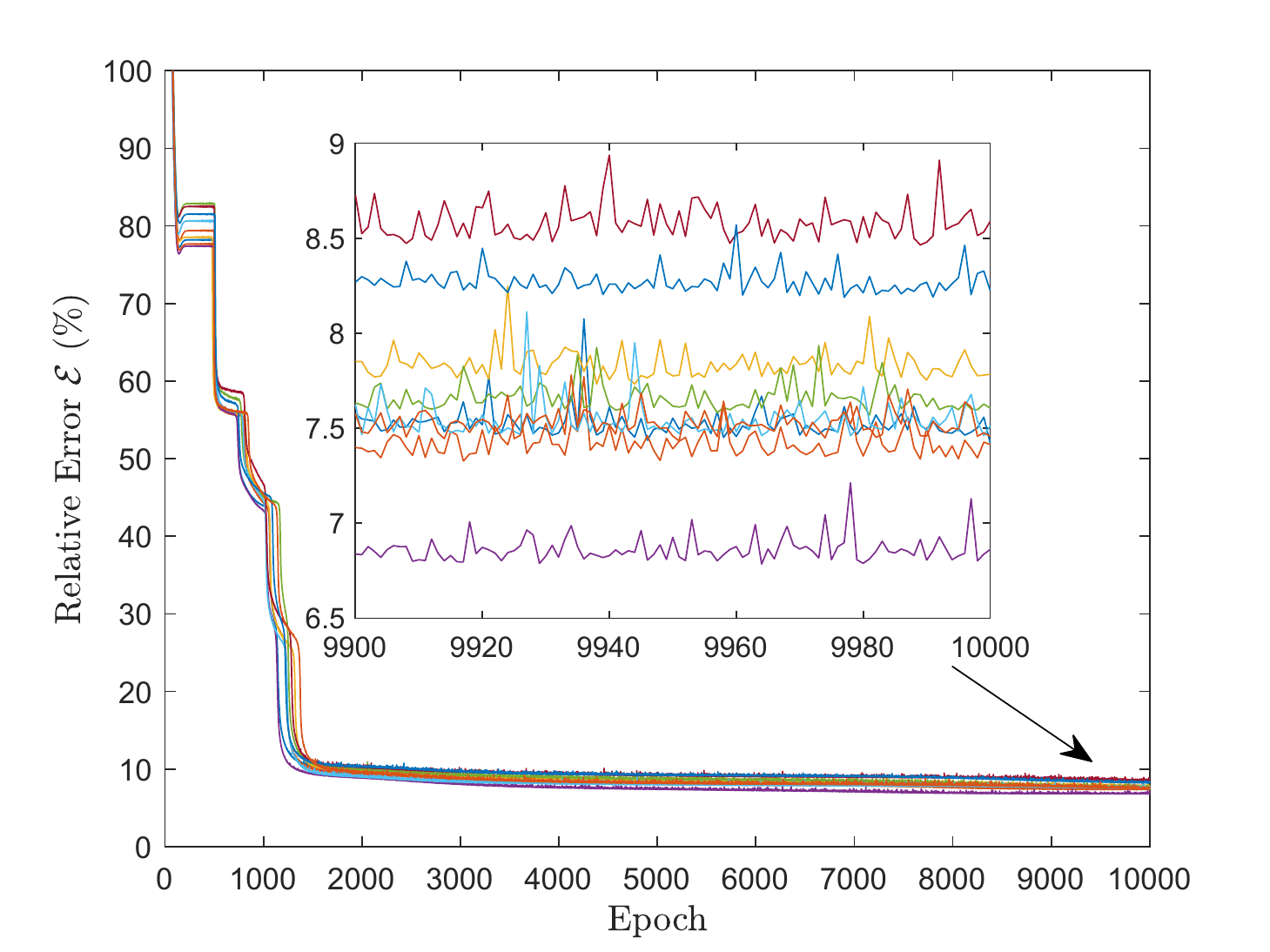}
\caption{Relative errors of nine independent training cases.}
\label{TANN2}
\end{figure}

\begin{figure}[!t]
\centering 
\includegraphics[width =2.8in]{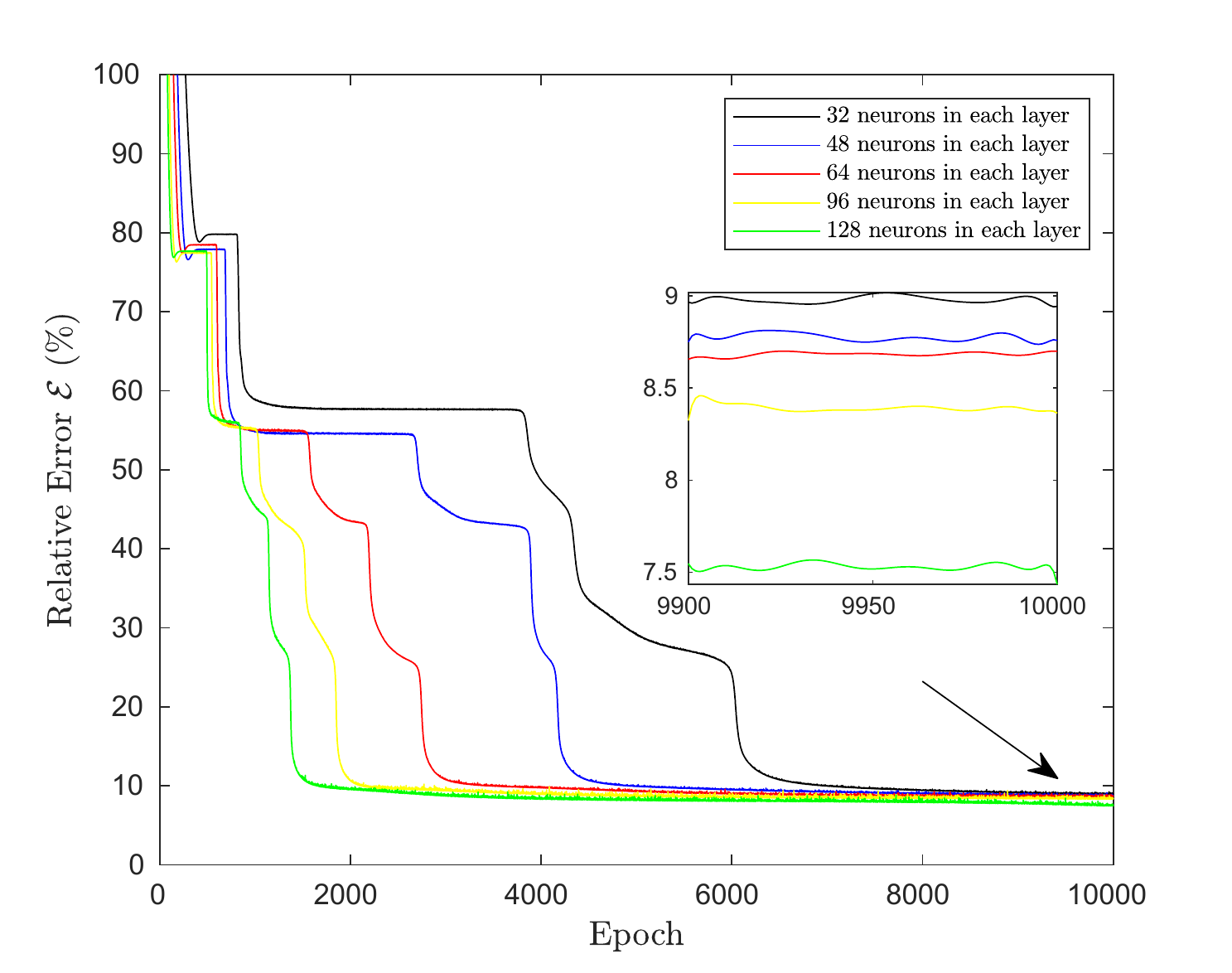}
\caption{Relative errors by different numbers of neurons in each layer.}
\label{neurons}
\end{figure}

\begin{figure}[!t]
\centering 
\includegraphics[width =2.8in]{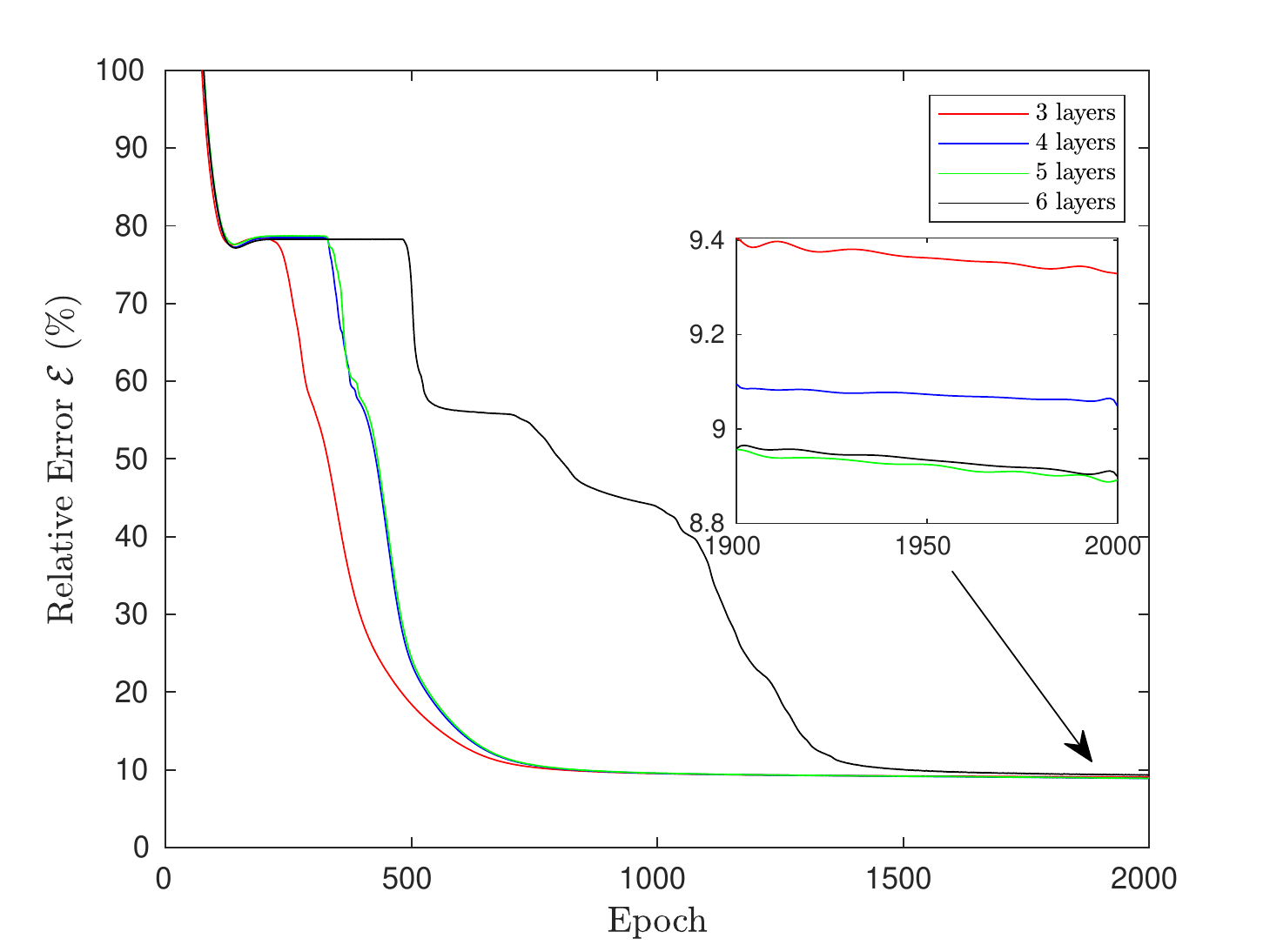}
\caption{Relative errors by different numbers of layers.}
\label{layers}
\end{figure}

\begin{figure}[!t]
\centering 
\includegraphics[width =2.8in]{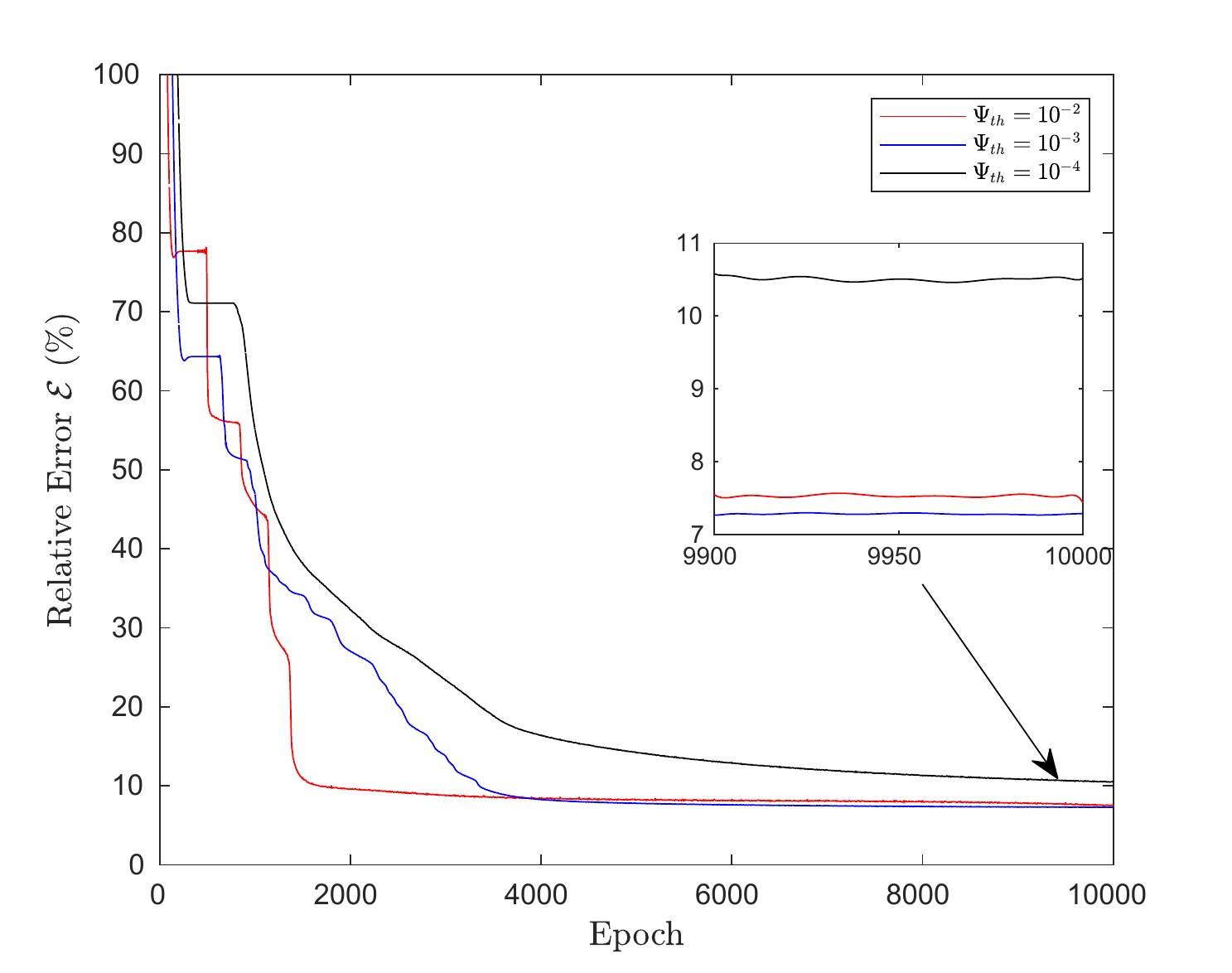}
\caption{Relative errors by different outage  thresholds.}
\label{outage}
\end{figure}

\begin{figure*}[t!]
    \begin{subfigure}[t]{0.3\textwidth}
        \includegraphics[width=2.3in]{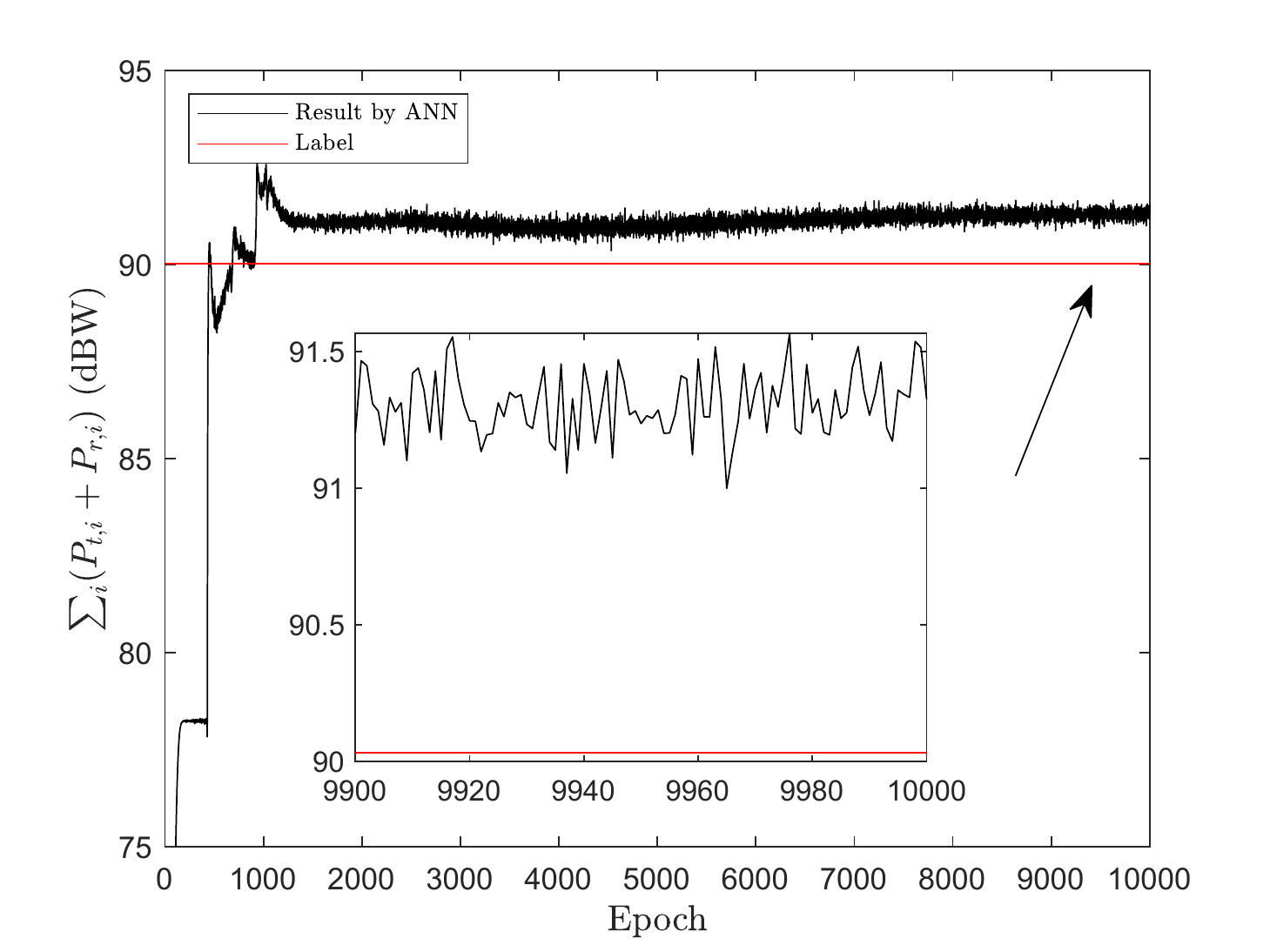}
        \caption{Example 1}
    \end{subfigure}
    ~~ 
    \begin{subfigure}[t]{0.3\textwidth}
        \includegraphics[width=2.3in]{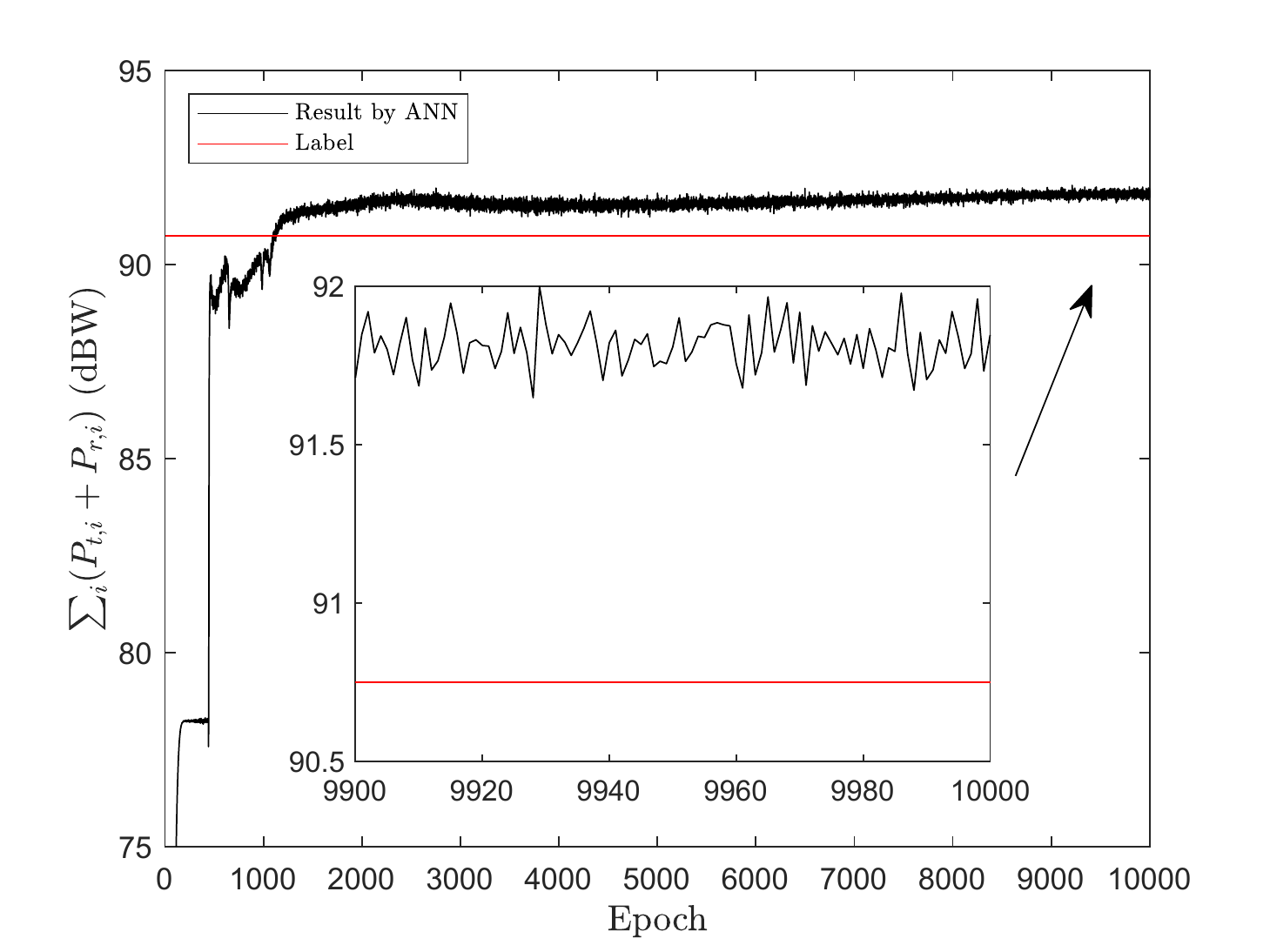}
        \caption{Example 2}
    \end{subfigure}
    ~~ 
    \begin{subfigure}[t]{0.3\textwidth}
        \includegraphics[width=2.3in]{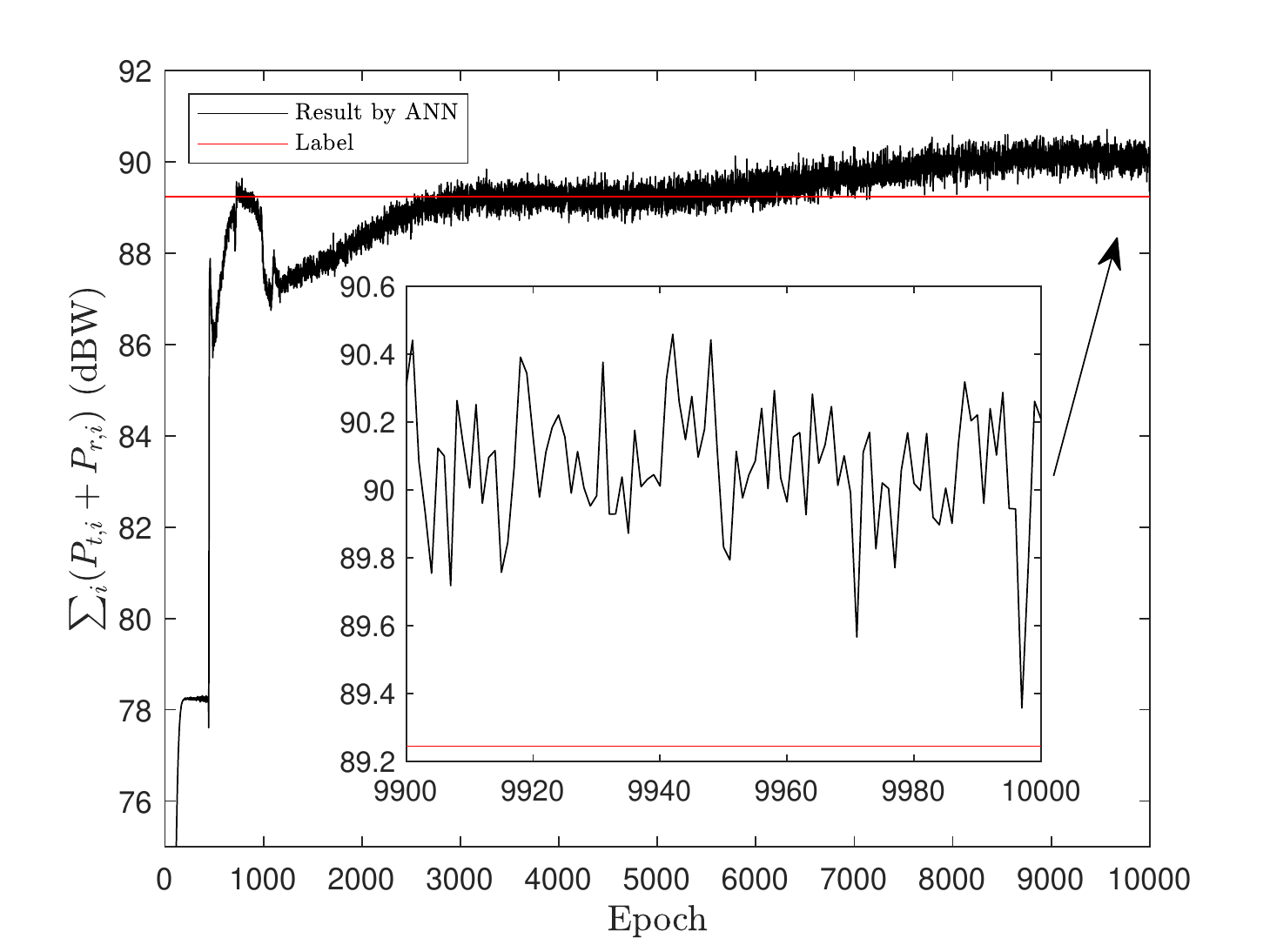}
        \caption{Example 3}
    \end{subfigure}
    \caption{Total transmit power of three independent cases yielded by the proposed ANN approach and exhaustive search.}
    \label{annplot}
\end{figure*}

We mainly compare the differences between the proposed ANN approach and the exhaustive search method.  In addition, considering the over-fitting hazard and the algorithmic generality, we generate another $L=1000$ groups of samples to test the accuracy of the trained ANN model. Although the loss function defined in (\ref{lossfunctioneq}) is appropriate for training purposes, it might not be quantitatively comparable for different groups of samples. For clarity, we introduce the relative error $\mathcal{E}$ as a metric for illustration purposes: $\mathcal{E} = \frac{1}{L}\sum_{l=1}^{L}\left(\frac{1}{2T}\left\|{\mathsf{abs}\left({\mathbf{\hat{U}}_{l}}-\Omega(\mathbf{\hat{V}}_{l},\{\theta_*\})\right)}./{\mathbf{\hat{U}}_{l}}\right\|_1 \right)$, where $\mathsf{abs}(\cdot)$ returns a matrix/vector with the same dimension as the argument and the absolute values of the matrix/vector entries; $./$ is the right array division that divides each entry of the dividend by the corresponding entry of the divisor; ${\mathbf{\hat{V}}_{l}}$ and ${\mathbf{\hat{U}}_{l}}$ denote the $l$th group of validation samples and the corresponding label. We plot the relative errors of nine independent cases in Fig.~\ref{TANN2}. It can be seen in Fig.~\ref{TANN2} that the relative error $\mathcal{E}$ gets lower with the increase of training epoch and approaches 7.5\%  which is acceptable for most practical applications. Moreover, by demonstrating the training performance for nine cases with different samples, the generality of the proposed power allocation scheme based on ANN and deep learning can be validated. 

Focusing on ANN architecture itself, we also provide some simulations for comparison purposes by varying the number of layers and the number of neurons in each layer. To reveal the statistical nature by the law of large numbers, we average the relative errors for all cases by 1000 repeated trials with different channel realizations to produce smooth curves. As shown in Fig.~\ref{neurons}, when we fix the number of layers to be six, average training efficiency has been obviously increased with an increased number of neurons in each layer. We demonstrate the effects of the number of layers on the average training performance in Fig.~\ref{layers} by fixing the number of neurons in each layer to be 128. Surprisingly, a smaller number of layers brings an even higher training efficiency at the beginning, whereas the relation reverses as expected when converging toward training limits.

We also vary the outage constraint in the range from $10^{-4}$ to $10^{-2}$ to study its impact on the average relative error. Fig.~\ref{outage} demonstrates that a higher outage constraint generally leads to a faster converging process. However, there is not a necessarily monotone relation between the outage constraint and the relative error when converging toward training limits. 

Meanwhile, as the optimization objective of the proposed optimization problem, total transmit power is also studied and simulated. We demonstrate the total transmit power of three independent cases in Fig.~\ref{annplot}. From this figure, we can observe that the proposed power allocation scheme is capable of yielding near-optimal performance compared to the exhaustive search method. In addition, by employing ANN, the computational complexity for power allocation has been greatly reduced. Consequently, the effectiveness and efficiency of the proposed power allocation scheme are corroborated by the numerical results, and the proposed power allocation scheme is numerically shown to be able to realize energy-efficient relayed OFDM-IM in real time with low complexity.

\bibliographystyle{IEEEtran}
\bibliography{bib}

\end{document}